# Locability: An Ability-Based Ranking Model for Virtual Reality Locomotion Techniques

RACHEL L. FRANZ

Computational Media and Arts, Hong Kong University of Science and Technology (Guangzhou), Guangzhou, Guangdong, China, rachelfranz@hkust-gz.edu.cn

JACOB O. WOBBROCK

The Information School | DUB Group, University of Washington, Seattle, Washington, USA, wobbrock@uw.edu

There are over a hundred virtual reality (VR) locomotion techniques that exist today, with new ones being designed as VR technology evolves. The different ways of controlling locomotion techniques (e.g., gestures, button inputs, body movements), along with the diversity of upper-body motor impairments, can make it difficult for a user to know which locomotion technique is best suited to their particular abilities. Moreover, trial-and-error can be difficult, time-consuming, and costly. Using machine learning techniques and data from 20 people with and without upper-body motor impairments, we developed a modeling approach to predict a ranked list of a user's fastest techniques based on questionnaire and interaction data. We found that a user's fastest technique could be predicted based on interaction data with 92% accuracy and that predicted locomotion times were within 12% of observed times. The model we trained could also rank six locomotion techniques based on speed with 61% accuracy and that predictions were within 8% of observed times. Our findings contribute to growing research in VR accessibility by taking an ability-based design approach to adapt systems to users' abilities.

## 1 INTRODUCTION[1]

The use of virtual reality (VR) is on the rise, and experts anticipate that the number of users will increase significantly—by almost 20%—within the next four years [24]. Once primarily associated with entertainment, VR's applications are expanding into multiple areas. It is now being used to connect with friends, family, and coworkers, offering a more immersive and realistic social experience compared to traditional video calls and other communication methods. Additionally, VR is increasingly used for creating art, visualizing data, engaging in sports, and participating in virtual travel experiences.

One of VR's most significant qualities is its ability to bring the world to the user, rather than requiring the user to go out into the world. This capability is particularly impactful for individuals with disabilities, who may face significant challenges traveling to see friends, work in physical office spaces, go to the gym, or explore new places. Given VR's potential to enhance the lives of so many people, it is crucial to ensure the technology is accessible to a broad range of users—especially those who stand to benefit the most, such as people with disabilities.

However, accessibility in VR poses unique challenges due to its inherently movement-based interaction paradigm. Navigating a virtual environment typically relies on *locomotion techniques*, which involve various forms of physical movement. Many of these techniques require upper-body mobility, such as reaching, swinging the arms, or turning the head. This reliance on physical motion makes VR less accessible to individuals with mobility impairments, particularly those affecting the upper body. Addressing these challenges is essential to make VR truly inclusive.

---

[1] Portions of this paper were published as Franz, R. L. (2024*). Supporting the Design, Selection, and Evaluation of Accessible Interaction Techniques for Virtual Reality*. (Publication No. 31490607) [Doctoral dissertation, University of Washington]. University of Washington ProQuest Dissertations & Theses.

Even though commercial VR is popular and novel, it has been the subject of research and development for several decades. Over this time, a wide range of locomotion techniques have been developed, utilizing different parts of the body, levels of physical exertion, and input modalities (e.g., 1D, 2D, 3D). These techniques require varying types of motor control, from fine motor skills to gross motor movements. To date, over one hundred distinct locomotion techniques have been documented, so many that researchers have created the "Locomotion Vault" (Di Luca et al., 2021).

The accessibility of these techniques depends on the particular impairment a user has, because individuals with the same medical diagnosis may experience different impairments that affect their ability to interact with VR. For instance, a person with a cervical spinal cord injury (SCI) might struggle with shoulder weakness, making it difficult to lift or reach with a controller. Conversely, another individual with SCI may have difficulty moving the thumbstick because they have limited fine motor control, but they are still able to lift and reach their controller. Thus, medical diagnosis is not a good proxy for understanding impairment.

For new VR users with impairments, identifying the most suitable locomotion technique can be daunting. Trying multiple techniques to find the best fit can be time-consuming, physically demanding, and even costly. A solution to this challenge is to have a user model that ranks optimal locomotion techniques based on a user's abilities, reducing the need for extensive trial-and-error. Such a system could begin with a quick questionnaire about the user's abilities to generate a personalized list of recommended techniques. Although filling out a questionnaire might be easy for the user, questionnaire data alone might not provide enough detail to make accurate predictions. Better predictions might be achieved by incorporating movement-related data collected during a calibration task. In this work, we explored building user models with these two approaches individually as well as a combination of the approaches.

We adopted an ability-based design approach [47, 48] to enhance VR accessibility through personalization, using machine learning techniques to develop predictive models for three scenarios. First, we describe the process for building and evaluating a predictive model based on questionnaire data about a user's abilities. Next, we describe the modeling process using movement- and performance-related data from a calibration task. Finally, we describe building and evaluating a model trained on a combination of questionnaire and calibration data. We found that the scenario in which the model is trained on calibration data only is the most efficient and effective. When ranking users' techniques, the model achieved an accuracy of 61%, with performance predictions within 8% of the observed performance. In addition, the model predicted users' fastest techniques with 92% accuracy, giving predictions within 12% of observed values.

This work makes two contributions: (1) A modeling approach to predicting a ranked list of a VR user's optimal locomotion techniques using questionnaire and/or calibration data, and (2) a model capable of predicting the fastest VR locomotion technique from a set of six options with high accuracy, as well as generating a ranked list of the six techniques with moderate accuracy, applicable to users both with and without impairments.

## 2 RELATED WORK

The research community is gradually making virtual reality (VR) more accessible for people with physical impairments; however, VR is still far from being accessible overall. In our work, we model user behavior to match users with techniques that could be accessible based on their individual abilities. Prior work related to our efforts can be found in accessible VR, ability-based design, and user modeling, each of which is addressed below.

**2.1 Accessibility in Virtual Reality**

As VR becomes mainstream, researchers have focused on improving its accessibility for people with disabilities, including people with vision, hearing, and mobility impairments. Some communities of practice, such as XR Access,[2] are working diligently toward identifying and addressing accessibility barriers endemic to existing VR systems. However, despite progress, VR hardware and interaction designs assume a typical body that is able to see, hear, stand, walk, reach, and crouch, making it inaccessible to people whose bodies do not conform to this standard.

VR shares many visual and auditory aspects with other technology paradigms, which hypothetically makes it straightforward to port assistive solutions that improve visual and auditory accessibility from other technologies to VR. Several studies have been conducted to improve the visual [6, 7, 16, 53, 54] and auditory [22, 31] accessibility of VR. However, in an attempt to convince the user that they are physically existing in a virtual space, VR uses a 3D interaction paradigm that does not exist for touch-based or desktop computing paradigms. As a result, initial exploratory research has been critical for understanding physical accessibility challenges for people who use wheelchairs and people who have upper-body impairments.

*2.1.1 Physical Impairments: Identifying Accessibility Challenges*

People with limited mobility identified several accessibility challenges using a commercial VR system including putting on and taking off the headset, using the buttons, and maintaining a view of their virtual controllers while wearing the headset (Mott et al. 2020). Baker et al. [1] performed an exploratory study of VR accessibility, which focused on the potential for VR to be used in long-term care facilities. Because all participants in the study used a wheelchair, a significant challenge was reaching virtual objects around the wheelchair without bumping into the armrests.

A survey by the Disability Visibility Project [49] found that individuals with self-reported disabilities who had used VR experienced several challenges. One challenge was that the motion tracking system assumed certain abilities (e.g., the ability to stand or use two hands). Another issue was that some respondents were unable to use motion as input because of differences in dexterity, reflexes, and range of motion. Respondents who used wheelchairs had difficulty moving or rotating their wheelchairs to the position expected by the VR system. To overcome these challenges, some respondents tried to use other hardware, such as game pads; however, most VR systems were "locked" and did not allow the use of other input devices.

Reinforcing these findings, Gerling and Spiel [14] identified ableist assumptions in the design of VR systems and concluded that the design does not accommodate bodies that do not adhere to an ideal standard. They suggested that designers should consider the "actions lent," or the actions a user must employ, to use a VR app's interaction paradigm and hardware. In this way, VR designers and researchers can identify physically inaccessible actions [14].

*2.1.2 Physical Impairments: Improving Accessibility for People with Upper-Body Impairments*

As a result of the above findings, researchers have worked to create systems and techniques that improve the accessibility of VR for people with upper-body impairments. Yamagami et al. [50] developed a framework that helps designers translate bimanual into unimanual interactions for people who only have use of one hand. Franz et al.

---

[2] https://xraccess.org/

[11] devised a taxonomy to help designers and researchers choose scene-viewing techniques that do not require head movement based on the characteristics of the user's virtual environment (VE). They also developed a framework to guide the creation of accessible point-of-interest techniques [10]. Designers could use the framework to create scene-viewing techniques that enabled users to view points-of-interest in the environment without having to turn their heads or bodies. To explore how users with spinal muscular atrophy would want to interact with VR, Tian et al. [45] elicited a set of gestures and found that users preferred to map large movements (i.e., reaching) into more localized gestures. They also found that users preferred gestures that could adapt to their changes in abilities over time. Along with general interaction techniques, an essential part of the experience of VR is being able to navigate within a VE. Despite the numerous locomotion techniques that exist, the accessibility of these techniques has yet to be investigated

*2.1.3 Accessibility of Locomotion Techniques*

The design space of locomotion techniques in VR is rich and varied. The simplest and most intuitive method to move in a VE is room-scale walking, in which the user walks around a physical space to move in a virtual space [46]. However, physical constraints, such as the size of the room, have resulted in researchers and designers inventing locomotion techniques that do not require a one-to-one mapping of physical to virtual movement [19, 26, 28, 29, 41]. In addition, walking techniques can require high physical exertion to perform. Designers have focused on controller-based locomotion techniques as well as stationary, movement-based techniques to address these problems. These techniques also have the potential to be used as accessible alternatives to walking-based techniques. Di Luca et al. [27] surveyed existing locomotion techniques in research and commercial applications. They found that room-scale and movement-based locomotion techniques tended to be less accessible than controller-based techniques. Franz et al. [12] found similar results, highlighting that *Teleport*, which is controlled by aiming the controller and pressing a button, *Astral Body*, which is operated using the controller's thumbstick, and *Sliding Looking*, with which users pressed a button and turned their heads, were the most accessible locomotion techniques for people with upper-body impairments. These findings reinforce the importance of offering a variety of locomotion techniques to accommodate users' abilities and preferences.

**2.2 Ability-Based Design**

Ability-based design (ABD) is an approach to accessible technology design that, unlike other approaches such as assistive technology design, argues that designers should leverage users' abilities rather than focusing on their disabilities, which can be achieved by designing systems that accommodate users [47, 48]. It also argues that systems should be scrutinized for their "ability assumptions," namely the abilities that they implicitly assume users have, and without which, users might not be successful. ABD systems should not, ideally, require assistive add-ons or peripheral devices to make technology accessible; rather, they should accommodate their users' abilities directly. This shift of focus also unburdens the user from needing the knowledge, finances, or time to acquire and learn how to use a specialized technology to make another technology accessible [48]. Instead, this burden is shifted to the technology. There is evidence to suggest that unburdening the user from technology acquisition and configuration could result in lower abandonment rates by people who use assistive technologies [2, 15]. VR systems could benefit from ABD by adapting to users' abilities, making off-the-shelf devices accessible. Our work adopts an ABD perspective by trying to understand a user's abilities, model those abilities, and provide the best matching locomotion techniques for those abilities.

### 2.3 Modeling User Behavior to Personalize Virtual Reality Interaction

A common method for personalizing technology involves modeling user behavior and adapting system components to better meet individual needs and abilities [39]. User modeling has been used to improve accessibility by optimizing desktop and mobile interfaces. And, although modeling has yet to be used to improve accessibility for VR users, it has been used to better understand user performance in VR and improve user authentication.

*2.3.1 Performance Modeling for Interface Optimization*

Several researchers have developed user models to improve touch accuracy and interface layouts on touchscreen devices for people with motor impairments [34, 36, 37, 42]. Mott et al. [36] modeled touch input by participants with motor impairments using template matching to predict participants' intended touch points, thereby reducing errors. In another study, Mott and Wobbrock [37] found that touch accuracy could be improved by predicting a touch point using a combination of individual and independent user models. Similarly, Montague et al. [34] employed shared user models to adjust target sizes and touch duration, finding that these adaptations significantly reduced errors for people with impairments. Gajos et al. [13] also took the approach of adapting interface elements and layouts to users' motor abilities, finding that users with motor impairments were more efficient and accurate when using a desktop interface adapted to their performance.

*2.3.2 Performance Modeling in Virtual Reality*

Because VR and augmented reality (AR) are relatively new technologies, little research has investigated modeling interactions in this context. Cabric et al. [5] extended the keystroke-level model (KLM) to a mixed reality context and modelled common gestures including controller button clicks, hand raising, mid-air tapping, and head pointing. Erazo and Pino [8] also used a KLM to predict users' performance time while using touchless hand gestures. These two models predicted performance with 3D gestures but did not factor in the impact of a user's abilities, which could potentially make this technology more accessible.

*2.3.3 Authentication in Virtual Reality*

Although more work has yet to be done on performance modeling in VR, researchers have measured and characterized movement in VR for the purpose of authenticating users. Studies have shown that authentication can happen in real time [32], across different VR systems (e.g., Oculus Quest, HTC Vive, and HTC Vive Cosmos) [33], with only head movement [43], and by performing different tasks [25, 40]. Pfeuffer et al. [40] proposed a set of measures to capture individual differences in movement based on body relation theories. They found that measures of the relationship between device pairs (e.g., distance between headset and right controller) were the most important features in models that identified users. Although not related to detecting movement differences between people with and without impairments, features that predict individual differences could help personalize VR experiences for people with a range of abilities.

The challenge of VR research is that it is not possible for past interaction paradigms (i.e., touch, desktop) to be directly transferred to the VR context. In addition, the appeal of VR is that it engages the body, so the restricted movement of past interaction paradigms would not be desirable anyway. This unique aspect, however, makes designing accessible experiences for people with physical impairments an acute challenge. One advantage of the VR interaction design space is that it is already well-populated, enriched by decades of research and applications. There is an opportunity to leverage this design space to adapt VR interaction to the abilities of people with impairments through user models.

## 3   USE CASE SCENARIOS

We discuss how a VR system could adapt to a user through three use-case scenarios, assuming the user is setting up their VR system for the first time. When first-time VR users engage with virtual reality, they are usually guided through a tutorial designed to teach them how to interact with the virtual environment (VE). This includes mastering controls for actions like selecting, grabbing, pointing, and navigating. However, the way controllers are used often varies significantly between different applications, leading to a lack of consistency in user interaction. For instance, one application might require users to point at an object with a laser-like ray to lift and move it, while another might have users interact by intersecting their virtual hand directly with the object.

Locomotion techniques for navigating in the VE also vary. For example, one application might use *Teleport*, where the user points at the ground with a beam emanating from their controller and jumps to that position. Another app might use *Arm Swinging*, where users swing their arms as they would when walking to move forward in the VE. Because the locomotion techniques vary between apps, some apps might be inaccessible for people with particular physical impairments. And physical impairments can manifest in many ways, affecting gross and fine motor control, which can make different locomotion techniques accessible or inaccessible depending on a users' particular impairments.

The vision for this work is that a new VR user could answer a questionnaire and/or perform a calibration task as a part of the VR set-up process. Then, based on this input, our system would recommend a list of locomotion techniques that are most suited to the user's physical abilities. For example, if the user has good fine motor control, the system might recommend thumbstick-controlled locomotion techniques. However, if a user has good gross motor control, the system might recommend movement-based techniques (e.g., *Arm Swinger*). As a result of this setup process, the user could employ their recommended locomotion techniques across apps, thus making those apps accessible, without requiring the user to try each one.[3]

There are three use cases that illustrate our vision for designing this hypothetical locomotion technique recommender system based on a user model. In all three use cases, a user with an upper-body impairment is using a VR system for the first time and would like to know which locomotion technique to use across apps. In the first scenario, the *questionnaire scenario (QS)*, the VR user responds to a questionnaire about their upper-body impairments. Using this data as input, the system then predicts a ranked list of optimal locomotion techniques for that user.

In the second scenario, the *calibration scenario (CS)*, a user calibrates the system by using a locomotion technique in a navigation task. The system records low-level input data from the controllers and headset, such as position, rotation, and velocity, and uses it as input for the model. The system then outputs a ranked list of locomotion techniques.

In the third scenario, the *questionnaire + calibration scenario (Q+CS)*, the user responds to a questionnaire about their impairments, similar to the initial step in the QS. Afterwards, the user performs a calibration task with a locomotion technique, like in the CS. Based on both sources of data, the system produces a ranked list of suitable locomotion techniques for that particular user.

---

[3] We recognize that apps themselves need to be configured to allow any number of possible locomotion techniques for our vision to become a reality. App architectures and locomotion techniques alike need to adopt interoperable standards. VR operating systems could also play an important role in addressing this challenge. In any case, these issues are beyond the current scope of this work.

In this work we explored how well each scenario performed when creating a ranked list of locomotion techniques for users with and without physical impairments. The next section describes the process for collecting and modeling the data.

## 4 METHOD

The goal of this study was to build models that could predict a ranked list of optimal locomotion techniques for a new VR user. Results from a study by Franz et al. [12] found that the optimal technique in terms performance varied for individual participants with impairments. *Teleport* was the fastest technique for most participants with impairments ($n$=14). However, the fastest technique was *Astral Body* for three participants, *Sliding Looking* for two participants, and *Chicken Acceleration* (controlled by head movement) for one participant. There was less variability in the group of participants without impairments: *Teleport* was the top technique for the majority of participants, but *Grab and Pull* was the top technique for one participant, and *Chicken Acceleration* for another. This variance, particularly for people with impairments, suggests that there is indeed a need to adapt to users' abilities. Therefore, we explored the use of questionnaire, performance, and low-level device data collected for one locomotion technique to produce a ranked list of the remaining locomotion techniques for each participant.

Building on these findings, the subsections below describe three different phases of conducting our current research: (1) *data collection*, during which data were collected from participants, (2) *data processing*, during which these data were transformed into features for the models, and (3) *model building*, during which these features were used to train machine learning models.

### 4.1 Data Collection

The data collection method for this study is based on prior work by Franz et al. [12]. This section provides highlights from that study most relevant to the current work.

#### 4.1.1 Participants

We recruited two types of participants: 20 participants with physical impairments and 20 participants without physical impairments. We recruited participants without impairments for this study so that the models could be trained on data representing a range of abilities. Eligibility criteria included being fluent in English, 18 years or older, and able to provide informed consent.

#### 4.1.2 Participants With Impairments

Twenty participants with self-reported upper-body motor impairments were recruited through a local advocacy group for people with motor impairments and the university's study recruitment advertising page. In addition to the above eligibility criteria, criteria for this group included that they were able to hold at least one Meta Quest 2 controller and wear the headset. The mean age was 39.4 years ($SD$=16.6). Self-reported medical conditions are listed in Table 1. In terms of VR use, 10 participants had never used VR, while the other 10 participants had used it in some capacity (e.g., with a smartphone).

Table 1. Participants' self-reported upper-body motor impairment conditions. From Franz et al. (2023).

| Participant | Condition |
|---|---|
| P00 | Left hand amputee |
| P01 | Spinal stenosis at 3C level |
| P02 | Ehler's Danlos Syndrome, Beals Syndrome (FBN2 gene mutation) causing progressive muscle weakness |
| P03 | C-5 quadriplegia |
| P04 | Peripheral neuropathy |
| P05 | Osteoarthritis, nerve damage |
| P06 | Hand tremor and weakness |
| P07 | Limb Girdle Muscular Dystrophy Type 2A |
| P08 | Cerebral palsy |
| P09 | Chronic joint pain, ongoing carpal tunnel syndrome |
| P10 | Paralysis, quadriplegia |
| P11 | Tetraplegia |
| P12 | Muscular dystrophy |
| P13 | Muscular dystrophy |
| P14 | C5-C6 incomplete spinal cord injury with functional quadriplegia |
| P15 | Nerve damage, severe muscle spasms, arthritis in neck, sacroiliac joint pain |
| P16 | Arthritis in both wrists, elbows, and shoulders from overuse, right arm fatigues easily |
| P17 | Peripheral neuropathy in all extremities |
| P18 | Transverse myelitis |
| P19 | Ehlers-Danlos Syndrome |

*4.1.3 Participants Without Impairments*

In addition to the eligibility criteria for both groups described above, criteria for the group of 20 participants without impairments included not having a permanent or temporary motor impairment that affected the upper body. Participants without impairments were recruited through word of mouth and the university's study recruitment advertising page. Within this group, 10 women and 10 men participated in the study. The mean age was 23.2 years (*SD*=5.0). Nine participants had never used VR, while the remaining 11 had used it with either a smartphone or headset.

*4.1.4 Apparatus*

For the study apparatus, we used a custom testbed that implemented six locomotion techniques. We also collected data with questionnaires.

**Custom Testbed:** A custom testbed was created using a Meta Quest 2 headset along with its controllers, which were connected to an Alienware m15 P79F laptop. The testbed was developed in C# using Unity 2019.4.17. The environment featured a horizontal plane placed in an otherwise empty VE. Six blue targets were positioned in a circular arrangement on the plane (Figure 1A).

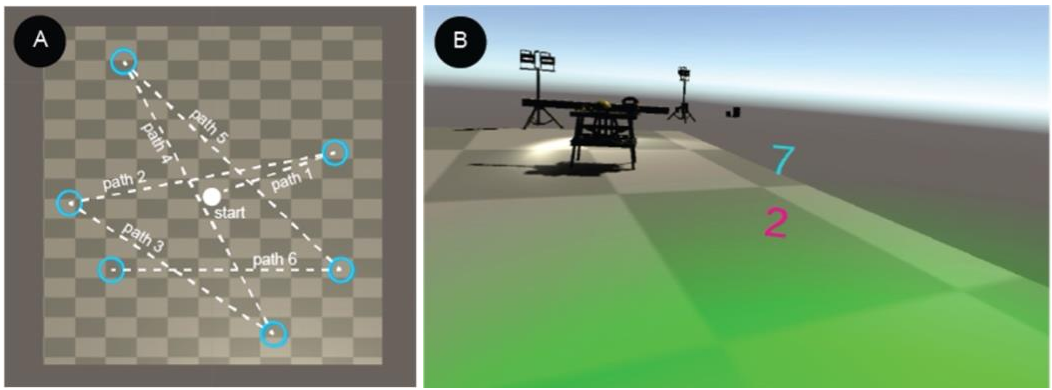

Figure 1. (A) A bird's-eye view of the test environment with the labeled path order and blue targets. (B) The participant's view when they moved into the active target, which appeared green. The blue countdown displayed the time remaining to complete the trial, while the pink countdown indicated the number of seconds the participant needed to stay within the target for it to be counted as a "hit." This design is similar to dwell-based selection on a touchscreen. Images from Franz et al. [12].

**Locomotion Techniques:** Six locomotion techniques were selected to ensure diversity in terms of the body parts used (e.g., head, torso, arms, fingers), control mechanisms (e.g., pointing, controller button manipulation, repetitive movement), and varying levels of effort (from low to high) (Table 2). Each technique could be used while seated (with no lower body movement) and those that required two hands could also be used with one. For all locomotion techniques except *Astral Body*, which employed a third-person perspective, participants could perform stationary turns using the thumbstick, which rotated the user's direction at 45-degree increments.

Table 2. The six locomotion techniques evaluated in this study, including usage instructions, the control mechanisms for both forward movement and direction, and the type of control mechanism. The table also details the body part involved, the effort level as defined by Di Luca et al. (2021), and the maximum number of hands required to perform the technique. Adapted from [12].

| Locomotion technique | Instructions | Forward movement | Direction | Control mechanism | Body part | Effort level | Hands |
|---|---|---|---|---|---|---|---|
| Astral Body | Use the thumbstick to move the avatar | Thumbstick | Thumbstick | Controller input | Thumb | low | 1 |
| Chicken Acceleration | Lean forward and look in the direction you want to move | Lean forward | Rotate head | Movement | Head and torso | medium | 0 |
| Grab and Pull | Reach forward, press and hold the trigger, pull the controller back toward you, then release the trigger | Reach and retract | Always forward | Movement, controller input | Arm(s), fingers | high | 2 |
| Sliding Looking | Press and hold the "A" or "X" button and look in the direction you want to move | Press the "A" or "X" button | Rotate head | Movement, controller input | Fingers, head | low | 1 |
| Teleport | Aim the blue circle and press trigger to jump to the circle | Aim and press trigger | Always forward | Aiming, controller input | Arm, finger | low | 1 |
| Throw Teleport | Press and hold the grip to show a ball. Throw the ball and release the grip to release ball. You will move to where the ball lands. | Aim and throw ball | Always forward | Movement, controller input | Arm(s), fingers | high | 2 |

**Self-Reported Impairment Questionnaires:** All 40 participants completed two questionnaires designed to assess the extent of upper-body impairments. The *Quick*DASH questionnaire [17] is a validated and shortened version of the 30-item Disabilities of the Arm, Shoulder, and Hand questionnaire (DASH) that measures physical function in the arm, shoulder, and hand [20]. We chose *Quick*DASH because it evaluates various aspects of upper-body function, does not depend on a specific medical diagnosis, and is concise, containing just 11 questions.[44]. Participants responded using a 5-point Likert scale, with values indicating increasing levels of difficulty. An example item included, "Please rate your ability to do the following activities in the last week by circling the number next to the appropriate response: Use a knife to cut food." The full questionnaire is available in Appendix A1. A composite score ranging from 0 to 100 was used to assess how musculoskeletal conditions affect upper-body function, with lower scores indicating more function. On average, participants with impairments scored 50.5 (*SD*=16.6), while those without impairments scored 4.6 (*SD*=3.8).

The second instrument was a custom-designed questionnaire informed by prior research in accessible computing. It synthesized characteristics of participants with motor impairments from studies by Mott et al. [35] and Findlater et al. [9] and were then reformatted into questions. This tool, referred to as the Technology-Related Physical Impairment Questionnaire (TRIQ), included 19 binary (yes/no) items assessing common impairments that might impact technology use. Sample items included "tremor," "poor coordination," and "slow movements." Among participants with impairments, the most frequently reported issues were "low strength in core, shoulders, neck, arms, hands, or fingers" (*N*=19), "limited wrist extension or flexion" (*N*=18), and "difficulty holding objects" (*N*=17). On average, these participants selected 5.8 impairments (*SD*=6.3). Appendix A2 contains the full questionnaire.

**Post-Task Questionnaire:** The post-task questionnaire included question #1 from Slater and Steed's (2000) presence questionnaire, which research has found to elicit the most direct response for presence. The first question from the Simulator Sickness Questionnaire, which measured general discomfort, was also used (Kennedy et al., 1993). Additionally, five questions were drawn from the NASA-TLX workload questionnaire (Hart & Staveland, 1988). For additional detail, see Franz et al. (2023).

*4.1.5  Procedure*

After signing the consent form, participants completed a demographic questionnaire along with the *Quick*DASH and TRIQ questionnaires. The researcher then loaded the test environment and guided the participants through practice sessions using the controllers and locomotion techniques. The participant could request to advance to the test phase once they felt comfortable with the controllers and techniques.

During the test phase, each participant used all six techniques. The test phase began with the participant positioned at the center of a circle of targets (Figure 1A). Once the initial countdown ended, the participant was rotated toward the first target, which turned green, and a 30-second countdown appeared in front of them, indicating the time remaining in the trial. Participants were instructed to move towards the target using one of the six locomotion techniques. Upon reaching the target, a five-second countdown appeared that indicated the time required for them to remain within the target area for the target to be considered "hit" (Figure 1B). If the 30-second countdown expired before the participant had remained inside the target for five seconds, the target was marked as "missed."

In cases in which participants did not reach the target before the 30-second countdown ended, the participants were automatically moved inside the missed target and rotated toward the next target. This process was repeated for all six targets, with each path between targets constituting one trial. Completing six trials made up one trial block. For two of the six paths, obstacles were introduced, and participants were instructed to navigate around them (Figure 2A). Participants completed two identical trial blocks with the assigned locomotion technique

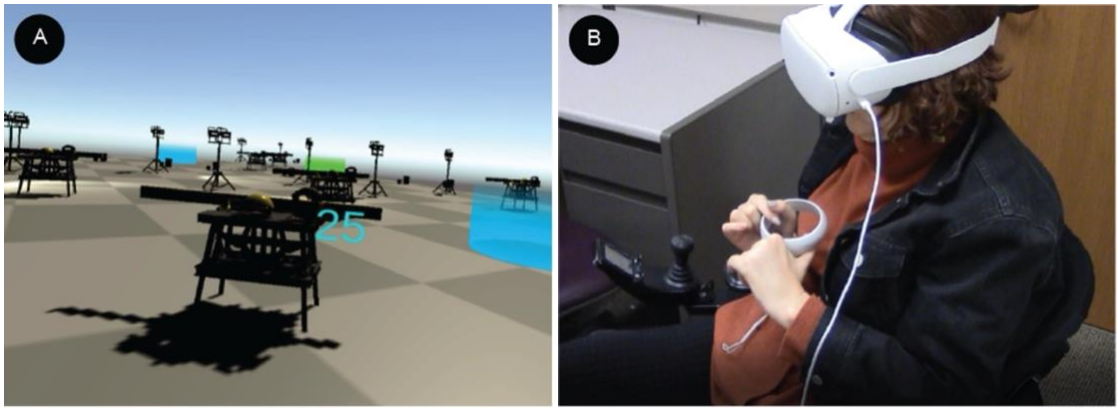

Figure 2. (A) The participant's view of the test environment, including the 30-second trial countdown. The target the participant is navigating toward is highlighted in green, while the other targets are blue. The participant must also maneuver around obstacles such as workbenches, lighting rigs, and paint cans. (B) A participant using one hand to stabilize the controller and the other to press the buttons while completing a trial. Image from Franz et al., 2023.

Following the test phase, the participant removed the headset and completed the post-task questionnaire. The researcher then asked the participant brief questions about their experience with the technique they just used. Following the interview, the participant put the headset back on and repeated the procedure for each subsequent technique. The order of the techniques was random.

Once the participant completed the procedure for all six locomotion techniques, the researcher conducted a final interview, asking the participant to compare all six techniques. The entire study session lasted between 2 and 2½ hours.

*4.1.6 Data Description*

Three types of data were collected during the study: Questionnaire, navigation task performance, and low-level controller and headset data.

**Questionnaire Data:** Questionnaire data consisted of *Quick*DASH responses to items and overall score, responses to individual TRIQ questions, and post-task questionnaire responses.

**Navigation Task Performance Measures:** There were three performance measures recorded per trial: *Trial Time* (in seconds), *Hit* (yes or no), and a count of the *Obstacles Hit*. Navigation task measures were important for representing a participants' success using the locomotion techniques.

**Low-Level Controller and Headset Data:** Low-level device data were collected from the motion controllers and virtual reality (VR) headset for the duration of each trial. These data included the *x*, *y*, and *z* values for position, rotation, velocity, angular velocity, acceleration, and angular acceleration of all three devices. Button data from the left and right controllers were also recorded and included thumbstick position, trigger and grip pressure, counts of trigger and grip presses, counts of primary and secondary button presses, and counts of primary and secondary button touches. Data also included the position of the controllers and headset within the VE, with the user's starting point at the origin. It would be this low-level controller and headset data that would be used in our user models, as described below.

## 4.2 Data Processing

This section describes the motivation for engineering features for our predictive model based on the low-level controller and headset data described above. Interview data revealed how participants believed their impairments influenced their use of locomotion techniques, but there was no empirical data to corroborate their verbal reports (Franz, 2023). Interviews and think-alouds are known to be limited when applied in user studies because people are often not able to verbalize their rationale for their actions [38]. This limitation could also have impacted participants' ability to identify how their performance with locomotion techniques was affected by their impairments.

By collecting low-level data from the headset and controllers as people with impairments use different locomotion techniques, we can better understand how their impairments affect their ability to control those techniques. Interaction-level metrics that capture how users' impairments influence their behavior might serve as valuable inputs for models predicting how well a new user will perform with specific locomotion techniques. Such metrics may also help detect upper-body impairments, improving the accuracy of performance predictions. As a result, we computed several metrics related to headset and controller use, which are available in Appendix A3.

## 4.3 Modeling

This section discusses the method used to create the predictive models for the questionnaire (QS), calibration (CS), and questionnaire + calibration (Q+CS) scenarios, and includes the target variables, datasets, feature selection method, machine learning algorithms, and model evaluation approach used. The QS is based on impairment-related questionnaire data. The CS is based on data from participants' use of one locomotion technique, which we refer to as the calibration technique. **The performance for the remaining five techniques was predicted using data from use of one technique to produce a ranked list of all six techniques**. The Q+CS is also based on data from participants' use of a calibration technique in addition to the same data used in the QS.

*4.3.1 Target Variable*

Three measures that captured different facets of performance were examined as the target variable (i.e., the variable being predicted). These measures included trial time, hit rate, and movement variability. Upon further exploration, it became evident that movement variability and hit rate were not good performance measures. There was a ceiling effect for hit rate across techniques: Most participants in both groups hit the targets. Movement variability did not seem to be affected by participants' impairments. Instead, it appeared to be influenced by the specific technique being used. Therefore, only trial time was used as the target variable.

In the QS, because there was only one set of responses at the participant level, the prediction was made at the technique level. In the CS and Q+CS, the prediction was made at the trial level.

*4.3.2 Model Features and Instances*

In the CS and Q+CS dataset, non-dominant-handed metrics were excluded because many participants with impairments did not use a controller with their non-dominant hand. Features for each scenario are presented in Table 3.

In the QS dataset, each instance represented one technique performed by one participant, for a maximum of 240 instances (40 participants × 6 techniques). However, one participant with impairments and two participants without impairments were excluded, for the following reasons: (1) The participant with impairments skipped over two questionnaire questions, (2) one participant without impairments did not finish all the techniques and (3) the other participant without impairments' controllers were not configured correctly, resulting in the final dataset containing 222 instances.

In the CS and Q+CS dataset, each instance represented one trial, so there was a maximum of 2,664 instances (6 techniques × 2 trial blocks × 6 trials × 37 participants). Some trials were missing due to technical issues during the study, so the final dataset contained 2,569 instances.

Table 3. The feature types, number of features, number of instances, and number of data points used for each scenario.

| | **Questionnaire Scenario (QS)** | **Calibration Scenario (CS)** | **Questionnaire + Calibration Scenario (Q+CS)** |
|---|---|---|---|
| Feature Types | 1. QuickDASH responses and score<br>2. TRIQ responses<br>3. Prediction technique name<br>4. Group (with or without impairment) | 1. Calibration technique movement-related metrics<br>2. Calibration technique button-related metrics<br>3. Calibration technique target-related metrics<br>4. Calibration technique performance measures (Trial Time, Hit Rate, Obstacles Hit)<br>5. Calibration technique post-task questionnaire responses (NASA-TLX)<br>6. Calibration technique name<br>7. Prediction technique name<br>8. Group (with or without impairment) | 1. QuickDASH responses and score<br>2. TRIQ responses<br>3. Calibration technique movement-related metrics<br>4. Calibration technique button-related metrics<br>5. Calibration technique target-related metrics<br>6. Calibration technique performance measures (Trial Time, Hit Rate, Obstacles Hit)<br>7. Calibration technique post-task questionnaire responses (NASA-TLX)<br>8. Calibration technique name<br>9. Prediction technique name<br>10. Group (with or without impairment) |
| Number of Features | 33 | 44 | 75 |
| Number of Instances: | 222 | 2,569 | 2,569 |
| Number of Data Points | 7,326 | 113,036 | 192,675 |

*4.3.3 Modeling Method*

Linear regression with elastic net regularization was used to model trial time in the QS due to the limited amount of data available. The simplicity of the algorithm coupled with regularization techniques prevented the model from overfitting the dataset. Elastic net regularization discourages reliance on any single feature or small sets of highly correlated features by penalizing large coefficients, making the model less sensitive to noise in the training data [52].

In the CS and Q+CS, random forest regression was used because of the larger dataset available compared to QS. Random forest regression can capture complex, nonlinear relationships between features and the target variable, and the averaging of multiple decision trees can minimize overfitting [4]. Trial times for five techniques were predicted using the data obtained while the participants were using the excluded technique (i.e., the calibration technique). Therefore, six models were built for both CS and Q+CS scenarios.

A grid search of tuning parameters was performed for each model, resulting in the highest performing models in terms of $R^2$. The `train` function in the R `caret`[4] package was used to train the models.

---

[4] https://cran.r-project.org/web/packages/caret/caret.pdf

*4.3.4 Feature Selection*

For all scenarios, feature selection was performed with `caret`'s implementation of recursive feature elimination (RFE) with linear regression. RFE, which is a wrapper-based feature selection technique was used because wrapper-based techniques are generally better compared to filter-based techniques at selecting the optimal features by removing correlated features [3]. Because there will likely be correlated features in the dataset, such as the *Quick*DASH score and question responses, a wrapper-based feature selection technique was appropriate.

For the CS and Q+CS, an optimal feature set was selected using recursive feature elimination (RFE) for each calibration technique. To avoid overfitting, only the top 20 features were used to train models if more than 20 features resulted from the RFE. Twenty was selected through trial-and-error as the number of features; this number is well below the 1:10 to 1:20 feature-to-instances ratio that is generally recommended for regression modeling [18].

*4.3.5 Model Evaluation*

Cross validation was performed with 25 to 27 folds, in which one or two participants were left out of the training set to estimate the performance of the model with unseen data. Folds were created using `caret`'s `groupKFold` function, which created training and testing sets based on participant ID. The average $R^2$ was computed across folds for an estimate of the model's accuracy. $R^2$ was selected as the performance metric because it is easily interpretable with values ranging from 0 to 1, where higher values equate to better performance. The $R^2$ value represents the amount of variability in the data that the model can explain, which provides insight into the model's goodness-of-fit. Root mean squared error (RMSE) was also recorded to capture the average magnitude of prediction errors; in other words, how much the predicted data deviated from the observed data.

*4.3.6 Ranking Techniques*

To rank techniques for each participant from best to worst, a ten-fold cross-validation procedure was conducted, and trial time was predicted for each hold-out set.

After extracting predictions from the ten-fold cross validation, we found the mean trial time for each technique per participant across the ten folds. Predicted and actual values were used to rank each technique from 1 to 6, where 1 represented the fastest technique and 6 represented the slowest technique. Rankings for actual and predicted trial times were recorded in a dataset, which also contained participant ID and the name of the locomotion technique. Mean absolute percent error (MAPE) was also calculated to assess how much the predictions deviated from the actual values on average and to facilitate comparisons between models.

## 5 RESULTS

In this section, we present findings from modeling trial time based on questionnaire responses, which we refer to as the questionnaire scenario (QS), data from a calibration task, which we refer to as the calibration scenario (CS), and a combination of calibration and questionnaire data, which we refer to as the questionnaire + calibration scenario (Q+CS).

### 5.1 Feature Selection

To train the models, we first found the optimal feature set for each scenario.

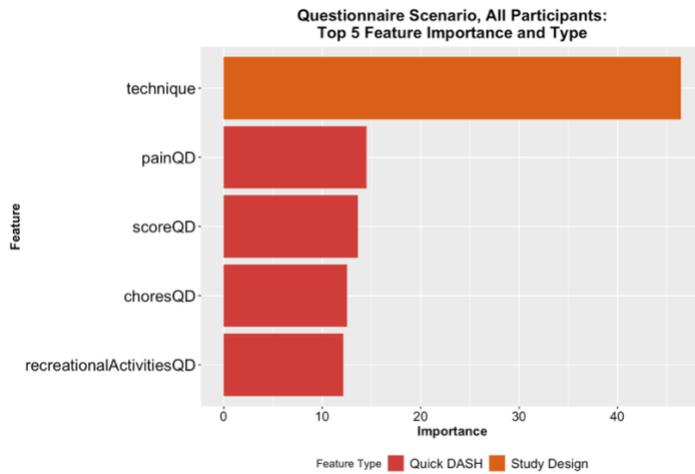

QS

Figure 3. A plot of the top five features for the questionnaire scenario (QS) model ordered by their importance ranking. Locomotion technique was highly ranked compared to all other features. The remaining features came from the *Quick*DASH questionnaire.

For the QS, a feature set of five was selected from the recursive feature elimination (RFE) process. The top features were all from the *Quick*DASH questionnaire, with the exception of prediction technique (i.e., the technique whose trial time was being predicted) (Figure 3). The prediction technique had an importance rating that was much higher than the other variables, which is reasonable given that some techniques took longer to use than others (e.g., *Grab and Pull* took longer than *Teleport* for all participants), so the trial time would largely depend on the technique being used. *PainQD*, which was an ordinal response to the statement, "On a scale of 1 (none) to 5 (extreme) rate the severity of your arm, shoulder, or hand pain in the last week," can also be explained as a top feature because several participants with impairments described feeling discomfort when using active techniques, such as *Grab and Pull* and *Throw Teleport*, regardless of the type of impairment they had [12]. Therefore, the speed at which participants could use the technique would likely depend on whether they were inhibited by discomfort or pain. Additionally, *scoreQD* can be explained as a top feature because it represented participants' overall impairment severity. *RecreationalActivitiesQD*, which is an ordinal response to the statement, "On a scale of 1 (no difficulty) to 5 (unable) rate your ability to do recreational activities in which you take some force or impact through your arm, shoulder or hand (e.g., golf, hammering, tennis, etc.) in the last week," might be explained as a top feature by the fact that recreational activities like sports would require similar capabilities as using some of the more active locomotion techniques such as *Grab and Pull* and *Throw Teleport*.

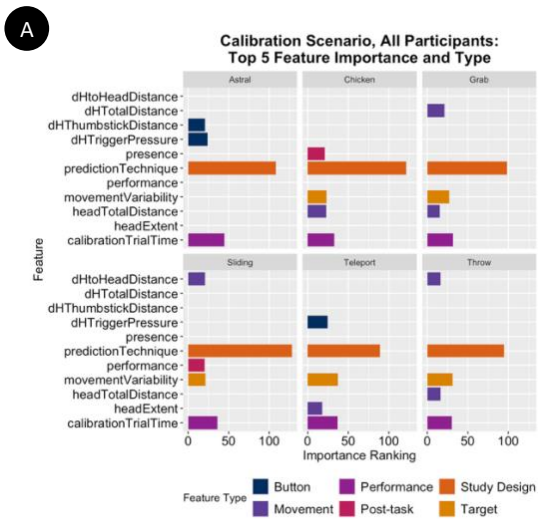

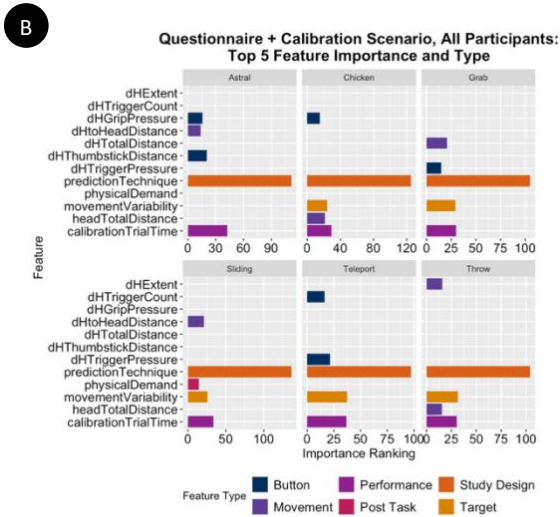

Figure 4. (A) The top five features for the calibration scenario (CS) and (B) questionnaire + calibration scenarios (Q+CS). "dH" represents "dominant hand". Each plot represents the top five features used by models to predict trial times, with the labeled technique serving as the calibration technique (the excluded technique). Besides the prediction technique (i.e., the technique whose trial time was being predicted), the trial time and movement variability for the calibration technique were the second- and third-most important features in most models.

In the CS and Q+CS, the prediction technique emerged as the most significant feature influencing trial time predictions (Figure 4). In addition, two other features played an important role: the time required to complete a trial with the calibration technique and movement variability (i.e., the extent to which users deviated from the optimal path) while using the calibration technique. The trial time for the calibration technique was a key factor likely because a participant who took longer to use the calibration technique would also likely take longer to use the prediction technique. Movement variability was likely important because it reflected participants' ability to control the techniques effectively, which would also impact the trial time.

Some of the top movement- and button- related features were also intuitive. For example, the total distance the head travelled was a top feature for *Chicken Acceleration* for both scenarios, which can be explained by the fact that *Chicken Acceleration* was controlled with the head. Dominant hand (DH) average trigger pressure was a top feature for *Teleport* in both scenarios, which aligns with the interaction mechanism because participants pushed the DH trigger button to select a location to move to. *Quick*DASH features were not among the top five features in either the CS or Q+CS models, even though they were top features in the QS model. This finding indicates that the questionnaire features were less influential on the model compared to calibration features. Three post-task questionnaire features, presence, performance, and physical demand, emerged as top features for *Sliding Looking* and *Chicken Acceleration* in both CS and Q+CS scenario models, indicating that group and individual differences could be explained in part by participants' perceptions of their use of the techniques.

*5.1.1 Model Training*

Upon training the model for the QS with the top features, we found that the model achieved an $R^2$ of 0.57 and a root mean squared error (RMSE) of 3.56 seconds (Table 4).

Table 4. $R^2$ values and root mean squared error (RMSE) for the model predicting *Trial Time* with the top questionnaire features in the questionnaire scenario (QS)

| QS | $R^2$ | 0.57 |
|---|---|---|
| | RMSE | 3.56 |

Upon training the models for the other two scenarios, the $R^2$ and RMSE values indicated that the best performing models across the CS and Q+CS were when *Astral Body* and *Chicken Acceleration* were used as calibration techniques in terms of both $R^2$ and RMSE (

Table 5). The other calibration techniques resulted in similarly performing models.

Table 5. $R^2$ and RMSE values for the calibration scenario (CS) and the questionnaire + calibration scenario (Q+CS) for each calibration technique, as well as the averages across calibration techniques. The values were similar for both scenarios.

| | | Calibration Technique | | | | | | |
|---|---|---|---|---|---|---|---|---|
| | | Astral | Chicken | Grab | Sliding | Teleport | Throw | Average |
| $R^2$ | CS | 0.56 | 0.65 | 0.33 | 0.35 | 0.34 | 0.34 | 0.43 |
| | Q+CS | 0.63 | 0.68 | 0.35 | 0.37 | 0.35 | 0.33 | 0.45 |
| RMSE | CS | 4.88 | 4.26 | 5.63 | 5.90 | 5.12 | 5.26 | 5.18 |
| | Q+CS | 4.47 | 4.06 | 5.36 | 5.60 | 5.07 | 5.28 | 4.97 |

Feature selection determined that the prediction technique was the most important feature in all three scenarios. This finding could indicate that trial time could be predicted with some accuracy by technique alone. However, training a model for trial time with prediction technique as the only feature resulted in an $R^2$ of only 0.37 and RMSE of 4.67 for the QS dataset, and an average of only $R^2$ 0.33 and RMSE of 6.02 for the CS and Q+CS dataset, indicating that questionnaire and calibration features improved model performance.

*5.1.2 Ranking Techniques*

Using the predicted times, we then ranked the techniques to observe how the models performed when recommending ranked lists of locomotion techniques to users. When ranking the techniques for the CS and Q+CS models, we selected the calibration technique with the models that had the lowest RMSEs, which were the *Chicken Acceleration* models. Our rationale was that these models would likely produce more accurate rankings compared to models with higher RMSEs because their predicted trial times would be closer to the observed trial times. Additionally, *Chicken Acceleration* would be a good calibration technique because it did not require any controller input, so it could be used by people without requiring use of their arms or hands.

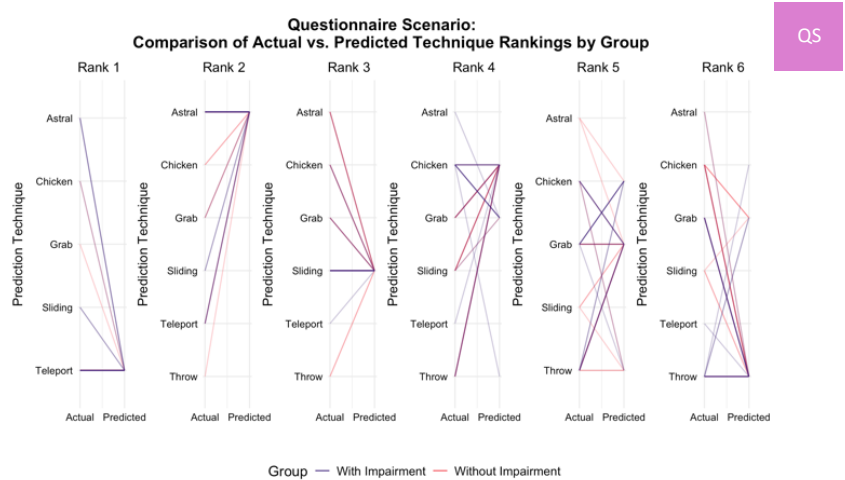

Figure 5. The QS slope graph of actual and predicted techniques by rank from fastest to slowest (rank 1-6). The *y*-axis is alphabetically ordered. Horizontal lines represent correct predictions and each line represents one participant. The questionnaire scenario (QS) model predicted the same technique within ranks 1, 2, and 3 for both groups, indicating that it was not able to model individual or group differences between self-reported impairment and trial time.

The QS model predicted the same technique within ranks 1-3 for both groups of participants, indicating that it was not able to model individual or group differences between self-reported impairment and trial time (Figure 5).

Thus, although the model was moderately accurate for predicting trial time, achieving an $R^2$ of 0.57 and RMSE of 3.56, it was mostly basing its estimates on the prediction technique, which is also what the feature selection results demonstrated. The fact that prediction technique alone could not explain as much of the variance of trial time as the final feature set, indicates that self-reported impairments *did* help the model predict trial time more accurately, but the model's relatively high performance when predicting trial time did not translate to a high accuracy in technique ranking.

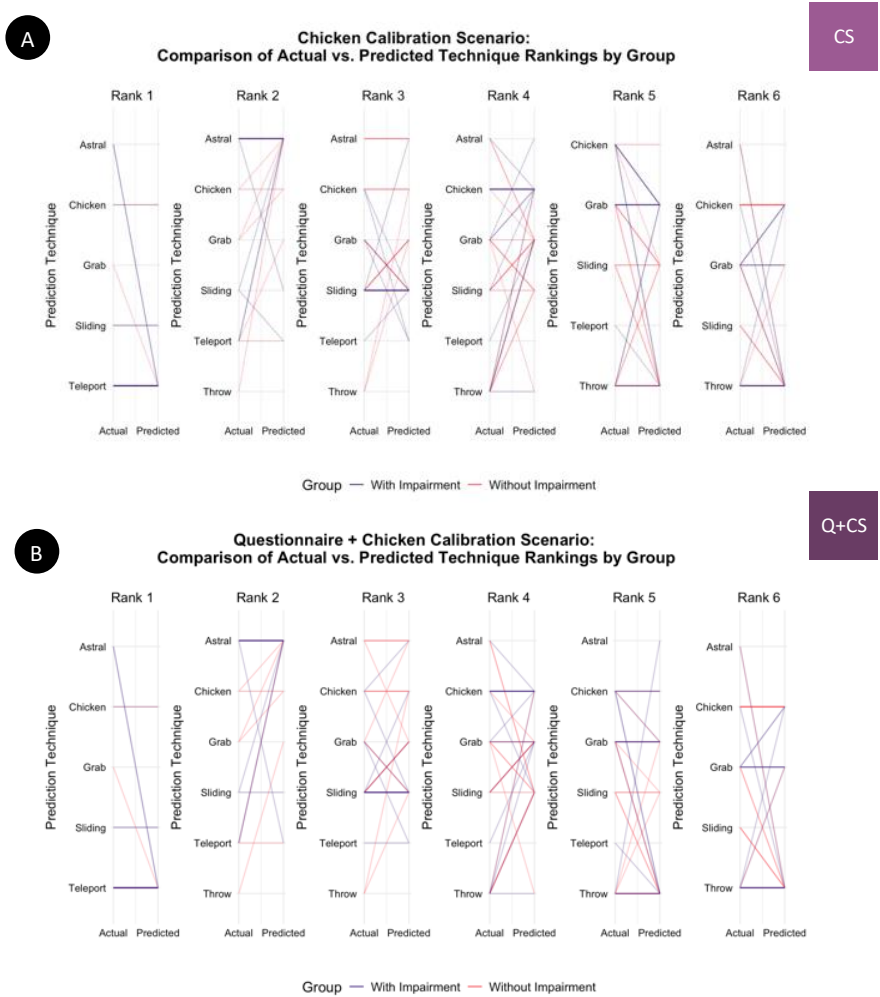

Figure 6. (A) Slope graph from the calibration scenario (CS). (B) Slope graph from the questionnaire + calibration scenario (Q+CS). Slope graphs show actual and predicted techniques by rank from fastest to slowest (ranks 1-6). The *y*-axis is alphabetically ordered. Horizontal lines represent correct predictions, and each line represents one participant.

The models for the CS and Q+CS with *Chicken Acceleration* as the calibration technique they were able to model some individual and group differences when ranking techniques (Figure 6). Some group differences were modeled for rank 6 in both scenarios as the models correctly predicted that *Grab and Pull* and *Throw Teleport* were the slowest techniques for people with impairments, while *Chicken Acceleration* was the slowest technique for people without impairments.

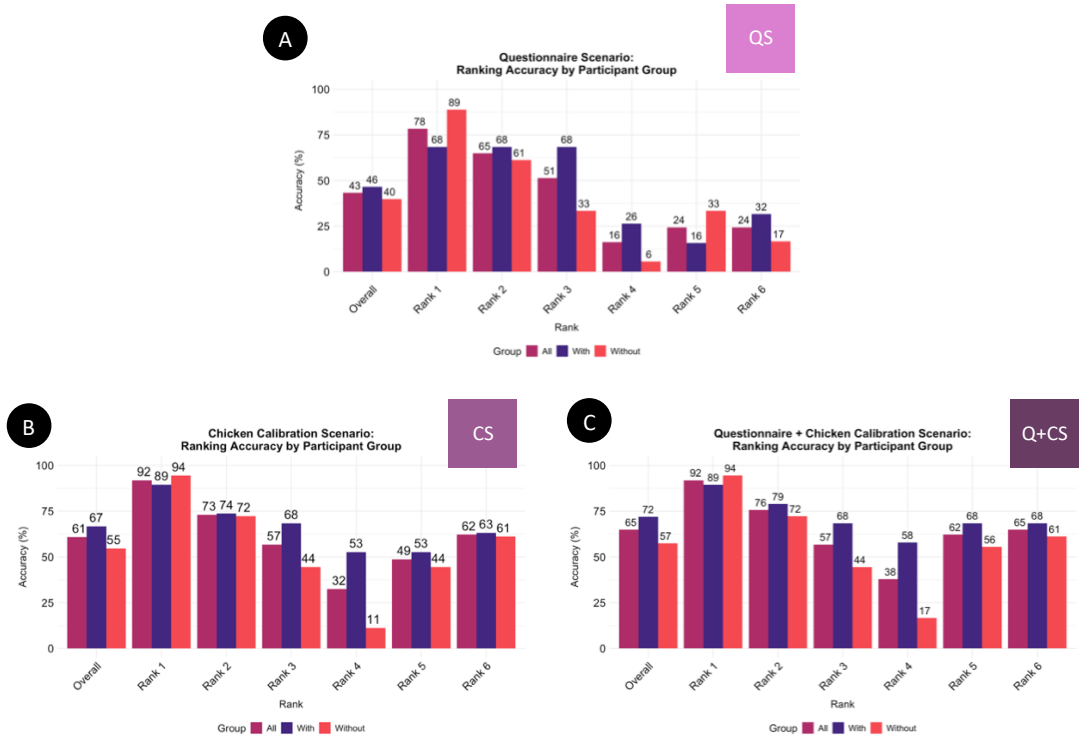

Figure 7. (A) Questionnaire scenario (QS), (B) Calibration scenario (CS), and (C) Questionnaire + calibration scenario (Q+CS) ranking accuracy by participant group. The CS and Q+CS models had the similar overall accuracy for all participants, which was higher than that for the QS.

Figure 7 illustrates that the QS had an overall lower ranking accuracy for all participants compared to the other two scenarios. The QS achieved an overall ranking accuracy of 43% while the CS achieved an accuracy of 61% and the Q+CS an accuracy of 65%. The Q+CS was slightly more accurate than the CS across ranks, and the two scenarios were similarly accurate when ranking the first technique, both achieving an accuracy of 92%. In the CS and Q+CS, people with and without impairments had similar ranking accuracies, with the exception of ranks 3 and 4, in which the models predicted the trial times more accurately for the group with impairments than the group without impairments. Across all three models, rank 1 had the highest overall accuracy and rank 4 had the lowest accuracy.

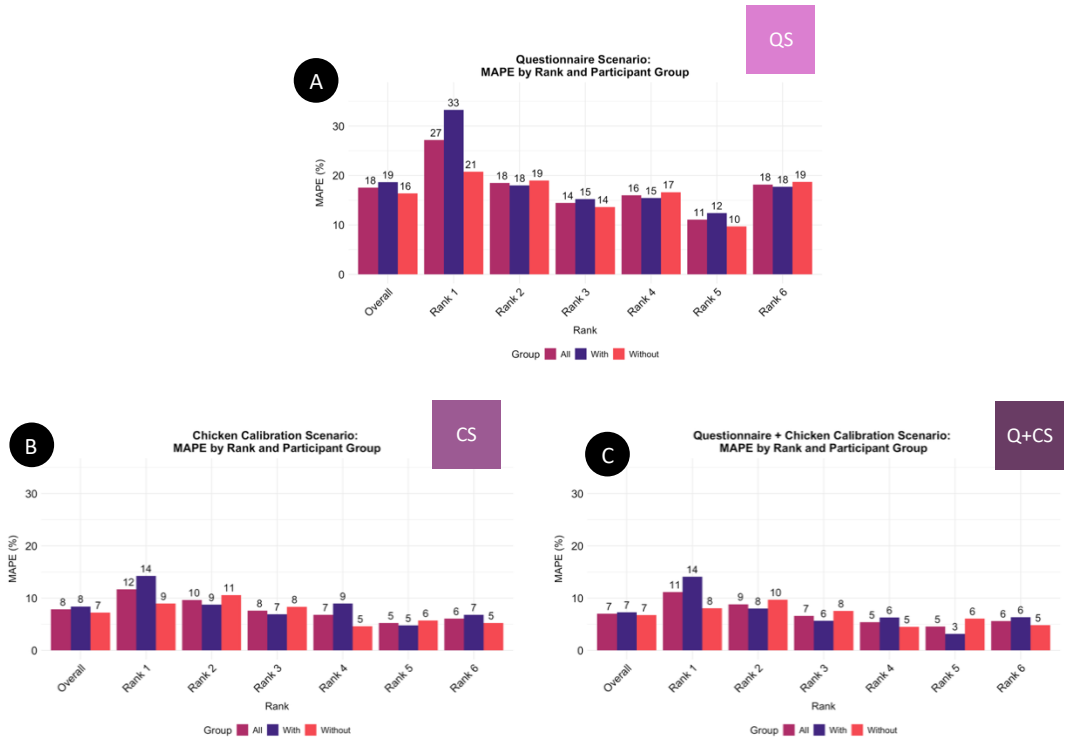

Figure 8. Left: The mean absolute error (MAPE) for the questionnaire scenario (QS) Middle: MAPE for the calibration scenario (CS). Right: MAPE for the questionnaire + calibration scenario (Q+CS). The CS and Q+CS had lower overall MAPE for all participants (lower MAPE is better). However, the CS and Q+CS models achieved particularly lower MAPE values for ranks three through six, indicating that they provided closer approximations of the observed trial times compared to the QS model.

As reflected in the ranking accuracy, the MAPE was highest for the QS while the CS and Q+CS had similarly low MAPE (lower is better). Overall, the QS predicted trial times that were within 17% of the actual trial times (Figure 8A), while the CS and Q+CS predicted trial times that were within 8% and 7% of the actual trial times, respectively (Figure 8B and C).

Even though the CS and Q+CS models performed worse on average in terms of $R^2$ and RMSE, they had a higher accuracy ranking and MAPE compared to the QS. This difference can be attributed to averaging trial times across the 12 trials in the dataset used for the CS and Q+CS, which brought the predicted average trial times closer to the observed average trial times. When the trial times were averaged, the $R^2$ for the CS and Q+CS when *Chicken Acceleration* was used for calibration was 0.89 and 0.90, respectively, compared to the initial $R^2$ of 0.65 and 0.68 respectively. Meanwhile, the averaged RMSEs were also lower for both scenarios: The RMSE was 1.88 *vs.* 4.26 for the CS and 1.80 vs. 4.06 for the Q+CS.

*5.1.3 Summary*

Modeling trial time with *Quick*DASH features resulted in a model that predicted participants' trial times moderately well. However, when ranking the predicted trial times and comparing them to actual trial times, the model predicted the same technique for the fastest three techniques as well as the slowest technique for each participant. By contrast, using calibration data in addition to questionnaire data, as well as calibration data alone, resulted in models that could rank participants' fastest techniques with 22% greater accuracy than questionnaire data alone.

## 6 DISCUSSION

We discuss our findings and reflect upon our modeling approach. We also offer three recommendations for creating user models in this domain.

### 6.1 Participant Impairments

Much of accessibility research focuses on populations with specific health conditions (e.g., spinal muscular atrophy). This is understandable because the assumption is that the same condition will impact peoples' technology use in a similar way. However, it is common for people with the same condition to have different types and severities of impairment.

This study used an ability-based design approach and focused on abilities that could be leveraged to adapt the system and make it accessible. As a result, our recruitment strategy was focused on what participants *could* do, namely hold a controller, push buttons, and wear a headset. The results from our study suggested that despite the variance in participants' diagnoses, the models could still find impairment-related use patterns in the features to train a high performing model, achieving 92% accuracy when predicting the fastest technique in the CS and Q+CS, making the model likely to be generalizable to individuals with the abilities specified in our inclusion criteria.

A potential research direction that could be explored is ability-based recruiting, in which participants are recruited based on what they can and cannot do relative to a specific technology. This could serve to broaden the ecological validity and transferability of study results across people with different conditions by grouping participants by technology-related impairment, rather than by diagnosis.

### 6.2 Locomotion Technique Ranking

As the QS revealed, predicting the fastest technique for most people for ranks 1 and 2, which were *Teleport* and *Astral Body*, still led to a relatively high accuracy and low MAPE: Rank 1 had an overall accuracy of 78% and a MAPE of 27%, while rank 2 had an overall accuracy of 65% and a MAPE of 18%. However, there was still enough variation within these ranks to justify personalizing the recommendation of techniques. Forcing a user to rely on a technique that does not align with their abilities can be discouraging, potentially leading to frustration and eventual abandonment of VR. By starting users with their optimal technique, the likelihood of successful adoption and continued use of VR can be improved.

The finding that *Teleport*, *Astral Body,* and *Sliding Looking* were the fastest for most people aligns with past research that found that techniques that do not require large movements tend to be favored by people who have fine motor control [51]. However, prior research has also found that people do not always want to use the fastest technique. Rather, qualities like visual appeal, entertainment, exercise, and rehabilitation were considered when participants chose their favorite techniques [12]. For example, Tian et al. [45] found that people with spinal muscular atrophy prefer VR interactions that that are visually appealing. Similarly, Franz et al. [10] found that participants preferred scene viewing techniques that matched the aesthetic of the virtual environment. In a different study, Franz et al. [12] found that several participants' favorite techniques differed from the ones they thought were most comfortable and easiest to use. Therefore, it is still important to produce a ranked list of techniques so that a user can select from a range of options with knowledge about how the technique might impact their speed. In the future, the model could also include user preferences as features, such as whether they want to experience more or less exertion, realism, and other relevant user experience factors to create a user model that is based on both abilities and preferences.

### 6.3 Questionnaire Scenario: Selected Features

In the questionnaire scenario (QS), the *Quick*DASH features had a higher importance ranking than TRIQ features during feature selection. *Quick*DASH is a standardized questionnaire, and questions have been designed to measure different dimensions of upper-body impairments [17]. TRIQ, on the other hand, has not been validated, so there are likely correlated questions that do not capture various facets of a person's impairments. Additionally, *Quick*DASH encoded more information about impairment severity by using a Likert-type scale, while TRIQ only encoded the presence of an impairment. TRIQ features are sparse because of their dichotomous nature, and many are likely redundant, which explains why *Quick*Dash features were selected over TRIQ features.

Even though *Quick*DASH had greater importance in the model than TRIQ features, a challenge with the *Quick*DASH features is that their meaning is not immediately evident. For example, it's not clear how difficulty doing chores relates to the ability to use *Chicken Acceleration*. This insight reveals an opportunity to create a standardized questionnaire that can capture technology-related impairments. As a result, a direct relationship could be drawn between the impairment and the effect on performance, facilitating quantitative studies that also explain results.

### 6.4 Calibration Scenario and Questionnaire + Calibration Scenario: Selected Features

The fact that the top features selected for each technique covered a range of data types (e.g., performance-related, movement-related, button-related, post-task, etc.) for both scenarios reveals that having a diversity of data types could have provided the model with a richer representation of an individual, leading to correct predictions of different techniques within ranks. Some of the button- and movement-related features had a clear relationship with the calibration technique. For example, a top feature for *Chicken Acceleration* was the total distance the headset travelled during a trial, which is intuitive because *Chicken Acceleration* is controlled by moving the head. Also, dominant hand (DH) extent was a top feature for *Throw Teleport* in the Q+CS, which is sensible because the user had to extend their dominant hand to throw the ball. These features serve as a sanity check to suggest that interaction with the techniques impacts performance, and the metrics we used likely capture differences in how users interact with the technique. In addition, a few post-task questionnaire items were also top features, suggesting that the way the participants perceived the technique also provided insight into their ability to use it.

Interestingly, there were no *Quick*DASH or TRIQ features selected as top features for the questionnaire + calibration scenario (Q+CS). This finding could be a result of a lack of variance in the questionnaire features. While the calibration data varied at the trial level, the questionnaire data only varied at the participant level. There may not have been enough variation for the model to pick up on individual and group-based patterns in the data to accurately predict trial times. This assumption is also supported by the fact that the model chose the fastest techniques for most people for ranks 1-3. However, despite not being in the top features, the questionnaire data improved the calibration model slightly as the Q+CS had a slightly higher accuracy and lower MAPE compared to the CS.

*Recommendation 1: Capture a variety of calibration data types.*
*When predicting trial times for the other techniques, the models relied upon a variety of data types as top features. These included how the user moved their controllers and pressed buttons, how they navigated within the space, and even how they perceived the calibration technique. Providing the model with various representations of users' abilities likely enabled it to identify underlying patterns between the features and trial times.*

### 6.5 Comparison of Scenarios

The findings from ranking the techniques indicated that the calibration scenario (CS) would lead to a more accurate ranked list of locomotion techniques compared to the questionnaire scenario (QS). The CS would also be preferable to the QS + CS because, even though the ranking accuracy and MAPE between the two scenarios were similar, the CS requires one less step because it does not require the user to fill out a questionnaire.

*Recommendation 2: Use the calibration scenario.*
*The calibration scenario led to higher accuracy in technique ranking compared to the questionnaire scenario and a similar accuracy to the questionnaire + calibration scenario. Since the calibration scenario does not require the user to fill out a questionnaire, it would be the most efficient and effective of the three scenarios.*

In terms of the calibration techniques, *Astral Body* and *Chicken Acceleration* models had similar $R^2$ and RMSE, therefore, either of these techniques might be used as a calibration technique. We chose *Chicken Acceleration* to create the ranked list because it had the lowest RMSE and did not require controller input. As a result, *Chicken Acceleration* might be accessible to a greater number of users during the calibration phase, including people who do not have use of their hands. *Astral Body* could also be a good calibration technique, as it only requires thumb movement.

*Recommendation 3: Use Chicken Acceleration or Astral Body as the calibration technique.*
*Chicken Acceleration and Astral Body performed better as calibration techniques in terms of $R^2$ and RMSE values compared to the other four techniques. Although we only evaluated the ranking accuracy and MAPE when Chicken Acceleration was used as the calibration technique, we expect that Astral Body would produce similar ranking results since its model had similar performance to that of Chicken Acceleration.*

### 6.6 Limitations and Future Work

A significant limitation of this predictive modeling approach is that the models would not be able to predict performance for a locomotion technique that was not used to train the model. If a designer was going to predict the trial time for a new technique, they would have to retrain the models with additional data. Without any examples of how people perform with a new technique to use as training data, it would not be possible to make a prediction with the modeling approach we proposed. In a future version of this work, one could imagine feeding a specification of a new technique into the system, a kind of description of what abilities it requires. Then, perhaps, the system could reason about its likely performance.

It also might be possible to estimate how a user would perform with a new technique by drawing comparisons to the techniques examined in the prediction study. For example, if the new technique uses gross arm movements like those in *Throw Teleport* and *Grab and Pull*, a designer could input the user's questionnaire and calibration data when they were using *Chicken Acceleration* to predict their trial times for *Throw Teleport* or *Grab and Pull* to help estimate their time for the new technique.

Adopting an alternative approach to creating a predictive model could be more effective in addressing the limitation of the current models only being able to predict the performance of techniques they have previously seen. For example, breaking down interactions into their component parts could help create a model that generalizes its predictions to new techniques. Interaction techniques can often be decomposed into discrete steps or "interaction units." For example, selecting a virtual target using a laser pointer in VR involves two primary interaction units: aiming the controller and pressing the trigger button. Each unit demands specific physical capabilities, such as arm strength, stability, grip strength, and finger dexterity. Performance predictions for new techniques can be made by developing a comprehensive vocabulary of interaction units common to several interaction techniques and mapping these units to the physical abilities required to perform them effectively. If associations between a user's abilities, interaction units, and performance with techniques that are composed of interaction units can be estimated, a predictive model could be built by mixing and matching interaction units. Predictions could then be made by plugging in the values for user abilities to produce a performance estimate without requiring training data for a specific technique.

## 7 CONCLUSION

This work demonstrates the potential to align users' physical abilities with the most suitable VR locomotion techniques. We introduce a user modeling approach that leverages low-level calibration data from VR devices to predict the fastest locomotion technique for each individual, regardless of whether they have a physical impairment. Furthermore, we developed a predictive model capable of accurately determining the optimal technique for a user from a set of six different locomotion techniques.

As the VR interaction landscape continues to expand, this ability-based approach offers a pathway to greater accessibility by matching user abilities to the interaction requirements of various techniques. This approach ensures that individuals with physical disabilities can use the same interaction techniques as those without disabilities while also selecting techniques that best suit their desired experience. By adopting this ability-based design framework, VR becomes not only more accessible for users with physical impairments but also more inclusive, fostering an equitable and engaging experience for users of all abilities.

## 8 ACKNOWLEDGEMENTS

This work was supported in part by a Facebook Social VR grant and by the Apple AI/ML Ph.D. Fellowship. Any opinions, findings, conclusions, or recommendations expressed in our work are those of the authors and do not necessarily reflect those of any supporter.

## A1. *QUICK*DASH QUESTIONNAIRE

### QuickDASH

Please rate your ability to do the following activities in the last week by circling the number below the appropriate response.

| | NO DIFFICULTY | MILD DIFFICULTY | MODERATE DIFFICULTY | SEVERE DIFFICULTY | UNABLE |
|---|---|---|---|---|---|
| 1. Open a tight or new jar. | 1 | 2 | 3 | 4 | 5 |
| 2. Do heavy household chores (e.g., wash walls, floors). | 1 | 2 | 3 | 4 | 5 |
| 3. Carry a shopping bag or briefcase. | 1 | 2 | 3 | 4 | 5 |
| 4. Wash your back. | 1 | 2 | 3 | 4 | 5 |
| 5. Use a knife to cut food. | 1 | 2 | 3 | 4 | 5 |
| 6. Recreational activities in which you take some force or impact through your arm, shoulder or hand (e.g., golf, hammering, tennis, etc.). | 1 | 2 | 3 | 4 | 5 |

| | NOT AT ALL | SLIGHTLY | MODERATELY | QUITE A BIT | EXTREMELY |
|---|---|---|---|---|---|
| 7. During the past week, *to what extent* has your arm, shoulder or hand problem interfered with your normal social activities with family, friends, neighbours or groups? | 1 | 2 | 3 | 4 | 5 |

| | NOT LIMITED AT ALL | SLIGHTLY LIMITED | MODERATELY LIMITED | VERY LIMITED | UNABLE |
|---|---|---|---|---|---|
| 8. During the past week, were you limited in your work or other regular daily activities as a result of your arm, shoulder or hand problem? | 1 | 2 | 3 | 4 | 5 |

Please rate the severity of the following symptoms in the last week. *(circle number)*

| | NONE | MILD | MODERATE | SEVERE | EXTREME |
|---|---|---|---|---|---|
| 9. Arm, shoulder or hand pain. | 1 | 2 | 3 | 4 | 5 |
| 10. Tingling (pins and needles) in your arm, shoulder or hand. | 1 | 2 | 3 | 4 | 5 |

| | NO DIFFICULTY | MILD DIFFICULTY | MODERATE DIFFICULTY | SEVERE DIFFICULTY | SO MUCH DIFFICULTY THAT I CAN'T SLEEP |
|---|---|---|---|---|---|
| 11. During the past week, how much difficulty have you had sleeping because of the pain in your arm, shoulder or hand? *(circle number)* | 1 | 2 | 3 | 4 | 5 |

$$\text{QuickDASH DISABILITY/SYMPTOM SCORE} = \left[ \left( \frac{\text{sum of n responses}}{n} \right) - 1 \right] \times 25,$$ where n is equal to the number of completed responses.

A *Quick*DASH score may **not** be calculated if there is greater than 1 missing item.

Figure 9.The *Quick*DASH questionnaire to evaluate upper-body impairment. This tool has been validated as a reliable assessment method [17].

## A2. TECHNOLOGY-RELATED (PHYSICAL) IMPAIRMENTS QUESTIONNAIRE (TRIQ)

Check all the impairments that you experience:

- Slow movements
- Low strength in core, shoulders, neck, arms, hands or fingers
- Tremor
- Poor coordination
- Rapid fatigue when using core, shoulders, neck, arms, hands or fingers
- Difficulty gripping objects
- Difficulty holding objects
- Lack of sensation in core, shoulders, neck, arms, hands or fingers

- Difficult controlling direction of hand or arm movement
- Difficulty controlling distance of hand or arm movement
- Limited range of motion in core, shoulders, neck, arms, hands, wrists or fingers
- Pain in core, shoulders, neck, arms, hands, wrists or fingers
- Poor precision with finger(s)
- Poor finger isolation
- Limited wrist extension or flexion
- Difficulty moving core, shoulders, neck, arms, hands or fingers quickly
- Difficulty moving core, shoulders, neck, arms, hands or fingers at the right time

- Difficulty balancing while seated
- Limited mobility of lower body (legs or feet)

Figure 10. The locomotion study utilized the **T**echnology-**R**elated (Physical) **I**mpairment **Q**uestionnaire (TRIQ), which is informed by literature in accessible computing but is not a standardized assessment tool. [9, 35].

**A3. ENGINEERED FEATURES[5]**

To identify features for predictive modeling, we designed movement-, button-, and target-based metrics, guided by earlier work examining interaction patterns among users with and without impairments (e.g., [21, 23, 30, 40]). Metrics were also shaped by insights from participant interviews conducted during a study comparing locomotion techniques with people with motor impairments [12]. We designed all metrics to be agnostic of any specific technique, which would allow for consistent comparison across different participant groups and interaction techniques.

We also hypothesized that certain metrics might be more relevant to specific techniques due to their control mechanisms. For example, head-related metrics were anticipated to influence trial time predictions for *Chicken Acceleration* and *Sliding Looking* techniques but not *Teleport*, because *Chicken Acceleration* and *Sliding Looking* are controlled with head movements. Observing differences between participant groups for some techniques but not others would support the notion that these differences were tied to how impairments affected interactions with specific techniques. A summary of the metrics analyzed in this study is provided in Table 6.

---

[5] This section is based on work that is currently under review for an ACM conference.

Table 6. The table outlines each proposed metric along with its description, classification, and rationale for why it may identify differences between participant groups. The term "device" encompasses the headset as well as the left and right controllers. All metrics are designed to be independent of specific techniques.

| Metric | Type | Description | Rationale |
|---|---|---|---|
| Device variability [23] | Movement-related | The total distance the device travelled | Could reflect the efficiency of movement |
| Device extent [23] | Movement-related | The Euclidean distance between the two furthest device positions | Could reflect the range of motion and efficiency of movement |
| Device angle variability [23] | Movement-related | The cumulative angular distance the device rotated | Could reflect the device stability |
| Device angle extent [23] | Movement-related | The angular distance between the two furthest angles | Could reflect the range of motion and device stability |
| Device velocity and acceleration | Movement-related | The average velocity and acceleration of the device | Could reflect the ability to control movement speed |
| Device angular velocity and acceleration | Movement-related | The average angular velocity and acceleration of the device | Could reflect the ability to control movement rotation speed |
| Total distance between device pairs [40] | Movement-related | Total Euclidean distance between device pairs (left-right, left-headset, right-headset) | Could reflect the efficiency of movement |
| Extent between device pairs [40] | Movement-related | Greatest Euclidean distance between device pairs | Could reflect range of motion |
| Number of submovements [21] | Movement-related | The number of movements between dips in velocity | Could reflect the efficiency of movement while performing the interaction |
| Thumbstick distance [23] | Button-related | The total distance the thumbstick traveled | Could reflect thumb strength and range of motion |
| Thumbstick extent [23] | Button-related | The angular distance between the two furthest angles | Could reflect thumb range of motion |
| Trigger and grip pressure | Button-related | Average trigger and grip pressure | Could reflect finger strength and control |
| Button press count | Button-related | Number of times a button was pressed | Could reflect finger dexterity and isolation |
| Target re-entry [30] | Target-related | Number of times the user exited and re-entered the target | Could reflect the ability to control speed, direction, and position |
| Target axis crossed count [30] | Target-related | Number of times the user crossed the optimal path to the target | Could reflect the ability to control speed, direction, and position |
| Movement variability [30] | Target-related | How straight the users' path is compared to the optimal path | Could reflect the ability to control speed, direction, and position |

**Movement-Related Metrics**

The device variability and (angular) extent metrics were adapted from prior work by Kong et al. [23], which focused on analyzing touch-based interaction data. Device variability might offer insight into the efficiency of user movements, while extent metrics could reflect both motion efficiency and the range of motion—such as torso movement for headsets and arm movement for controllers. Interview data reported by Franz et al. [12] indicate that these metrics might reveal group-level differences. For example, participants with impairments mentioned having difficulty turning their heads with techniques like *Chicken Acceleration* and *Sliding Looking*, which could impact headset-based angular metrics. Likewise, challenges with reaching and performing repeated motions during *Grab and Pull* and *Throw Teleport* techniques might be reflected in controller-based metrics such as variability, extent, velocity, and acceleration. Angular velocity and acceleration values were averaged across each trial, while the other metrics were computed using the formulas provided below.

$$Device\ Variability = \sum_{i=1}^{n-1} \sqrt{(x_i - x_{i-1})^2 + (y_i - y_{i-1})^2 + (z_i - z_{i-1})^2} \in [0, \infty)$$

$$Device\ Extent = \max_{i,j \in 0,\ldots,n-1} \sqrt{(x_j - x_i)^2 + (y_j - y_i)^2 + (z_j - z_{j-1})^2} \in [0, \infty)$$

Device Angular Variability and Extent were found in a similar way, where R represents the Euler angle:

$$Device\ Angular\ Variability = \sum_{i=1}^{n-1} \sqrt{(R_{ix} - R_{i-1x})^2 + (R_{iy} - R_{i-1y})^2 + (R_{iz} - R_{i-1z})^2} \in [0, \infty)$$

$$Device\ Angular\ Extent = \max_{i,j \in 0,\ldots,n-1} \sqrt{(R_{jx} - R_{ix})^2 + (R_{jy} - R_{iy})^2 + (R_{jz} - R_{iz})^2} \in [0, 360]$$

The device-pair variability and extent metrics are based on the work of Pfeuffer et al. [40], who investigated behavioral biometrics in VR. Their study examined motion-based features to capture individual movement patterns for the purpose of user authentication. Notably, device-pair metrics frequently emerged as key predictors during feature selection, underscoring their relevance. Due to their effectiveness in differentiating individual users, these metrics may also help distinguish between users with and without physical impairments. The calculations for these metrics are outlined below.

$$Device\ Pair\ Variability = \sum_{i=1}^{n-1} \sqrt{(x_{device1_i} - x_{device2_i})^2 + (y_{device1_i} - y_{device2_i})^2 + (z_{device1_i} - z_{device2_i})^2} \in [0, \infty)$$

$$Device\ Pair\ Extent = \max_{i,j \in 0,\ldots,n-1} \sqrt{(x_{device1_i} - x_{device2_i})^2 + (y_{device1_i} - y_{device2_i})^2 + (z_{device1_i} - z_{device2_i})^2} \in [0, \infty)$$

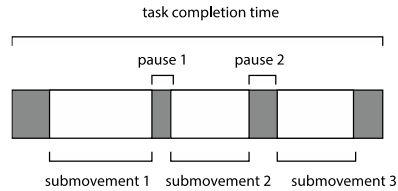

Figure 11. Submovements are movements within a rapid aimed movement with dips in velocity (pauses) before and after their occurrence. Figure adapted from [21].

The number of submovements metric is inspired by Hwang et al.'s [21] study, which examined movement patterns in motion-impaired individuals using a mouse. In their work, a submovement is defined as a segment of a rapid aimed motion where a drop in velocity occurs before and after the movement (Figure 11). The total number of submovements is calculated by summing all such instances within a trial.

**Button-Related Metrics**

The button-related metrics, while not derived from previous research, were designed to capture various user abilities such as finger isolation, control, range of motion, and strength. Metrics like thumbstick variability and extent have the potential to reflect these characteristics. During interviews, participants with impairments mentioned challenges with head movement, leading them to compensate by using the thumbstick for turning. This behavior would likely be reflected in thumbstick variability and extent. The formulas used to calculate thumbstick variability and extent are provided below.

$$Thumbstick\ Variability = \sum_{i=1}^{n-1} \sqrt{(x_i - x_{i-1})^2 + (y_i - y_{i-1})^2} \in [0, \infty)$$

$$Thumbstick\ Extent = \max_{i,j \in 0,\ldots,n-1} \sqrt{(x_j - x_i)^2 + (y_j - y_i)^2} \in [0, 1.2]$$

Participants with impairments noted difficulties pressing buttons and mentioned that their fingers occasionally slipped off them. These challenges could be reflected in grip and trigger pressure data as well as the number of button presses recorded. Average grip and trigger pressure were calculated by averaging pressure values across a trial, while button press counts were determined by summing the number of presses for triggers, grips, and primary or secondary buttons during a trial.

**Target-related Metrics**

Target-related metrics are adapted from MacKenzie et al.'s [30] framework for evaluating the accuracy of 2D pointing devices, modified here for use in a 3D environment. These metrics aim to capture a participant's ability to control position, direction, and speed with a locomotion technique by comparing their movement path to the optimal path—a straight line from the start to the target in the testbed environment (Figure 4).

During interviews, participants with impairments reported challenges with navigation, specifically noting poor control with *Chicken Acceleration* and *Throw Teleport* techniques, as well as difficulty determining if they were inside the target area with *Chicken Acceleration* and *Teleport*. These observations suggest that target-related metrics may help explain differences in performance between participants with and without impairments.

The first metric, movement variability (Figure 12), measures "the extent to which the sample points lie in a straight line along an axis parallel to the task axis" [30]. It is calculated using the following formula:

$$Movement\ variability = \sqrt{\frac{\sum_{i=1}^{n-1} y_i - \overline{y}}{n-1}}$$

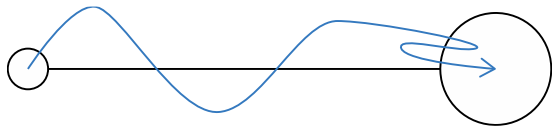

Figure 12. In this illustration, the user's path is represented by the blue line, while the optimal path is shown as a horizontal line. Movement variability quantifies how much the user's path deviates from the optimal path. The figure also shows an instance where the user first enters the target, exits it, and then re-enters. In this scenario, the target re-entry count would be recorded as 1. Additionally, the user is shown crossing the optimal path to the target twice, making the axis crossed count equal to 2.

The second metric, target re-entry count, is calculated as the total number of times the participant exits and re-enters the target (Figure 4). The third metric, axis crossed count, represents the number of times the participant's path intersects the optimal path to the target (Figure 4).

This paper extends a line of research initiated in a 2023 ASSETS publication[6], which reported on a study of six locomotion techniques for VR navigation, focusing on the performance and perceptions of users with motor impairments. The current manuscript builds on that work by introducing a predictive model trained on low-level device data (e.g. controller position), self-report questionnaires (about impairments), and performance metrics (e.g., trial completion time) collected during the same study. Unlike the ASSETS'23 paper, this work includes data from users with and without impairments.

Additionally, this manuscript incorporates interaction metrics explored in a separate paper currently under review (Appendix A3). That paper analyzes whether metrics such as total Euclidean headset distance differ significantly between people with and without impairments and how those differences might explain locomotion performance disparities. The current submission uses those metrics as features in our user model.

While these three papers use data collected during the same study, they offer distinct and complementary contributions: the ASSETS'23 paper contributes findings about the accessibility of six locomotion techniques, the paper under review investigates whether movement-based metrics can provide insight into locomotion performance, and the present submission proposes a modeling approach to recommend locomotion techniques based on users' abilities. Together, they form a cohesive research agenda documented in the author's dissertation[7].

---

[6] Rachel L. Franz, Jinghan Yu, and Jacob O. Wobbrock. 2023. Comparing locomotion techniques in virtual reality for people with upper-body motor impairments. In *Proceedings of the Conference on Computers and Accessibility (ASSETS'23)*, 2023. 1–15. https://doi.org/10.1145/3597638.3608394

[7] Franz, R. L. (2024*). Supporting the Design, Selection, and Evaluation of Accessible Interaction Techniques for Virtual Reality*. (Publication No. 31490607) [Doctoral dissertation, University of Washington]. University of Washington ProQuest Dissertations & Theses.